\documentclass[letterpaper, 10 pt, conference]{ieeeconf}  

\IEEEoverridecommandlockouts                              
\usepackage{algorithmic}
\usepackage[ruled,vlined]{algorithm2e}
\usepackage{amsmath,amssymb,amsfonts}
\usepackage{bm}
\usepackage{booktabs}
\usepackage{cite}
\usepackage{comment}
\usepackage{fancyhdr}
\usepackage{float}
\usepackage{fullpage}
\usepackage{gensymb}
\usepackage{graphicx}
\usepackage{hyperref}
\usepackage{mathrsfs}
\usepackage{textcomp}
\usepackage{times}
\usepackage{xcolor}
\usepackage{url}

\DeclareMathOperator*{\argmin}{argmin}

\begin{document}
\title{\LARGE \bf
Predictive Optimal Control with Data-Based Disturbance Scenario Tree Approximation}

\author{Ran Jing and Xiangrui Zeng 
\thanks{The authors are with the Department of Robotics Engineering, Worcester Polytechnic Institute,
        Worcester, MA 01609 USA
        {\tt\small \{rjing,xzeng2\}@wpi.edu}. Corresponding author: Xiangrui Zeng.}%
}

\maketitle
\pagestyle{empty}  
\thispagestyle{empty} 

\begin{abstract}
Efficiently computing the optimal control policy concerning a complicated future with stochastic disturbance has always been a challenge.
The predicted stochastic future disturbance can be represented by a scenario tree, but solving the optimal control problem with a scenario tree is usually computationally demanding. 
In this paper, we propose a data-based clustering approximation method for the scenario tree representation. 
Differently from the popular Markov chain approximation, the proposed method can retain information from previous steps while keeping the state space size small. 
Then the predictive optimal control problem can be approximately solved with reduced computational load using dynamic programming.
The proposed method is evaluated in numerical examples and compared with the method which considers the disturbance as a non-stationary Markov chain. 
The results show that the proposed method can achieve better control performance than the Markov chain method.
\end{abstract}


\section{Introduction}

Optimal control has been applied in many control applications for systems with uncertain disturbance. 
The disturbance considered in this paper can be generalized to any variables in the system dynamics equations which appear in the form of disturbance.
The actual physical meanings of these variables may be noise, external input, or environmental signals.
For example, the driver's power demand may be considered as generalized disturbance to the vehicle, and the ambient temperature may also be considered as generalized disturbance to the building heating, ventilation, and air conditioning systems.
Some disturbance may have high impacts on the optimal control cost, and the optimal control policies may depend on the characteristics of the disturbance \cite{zeng2019worst}.

In real-time optimal control problems, future disturbance information may be critical to the control policy.
Therefore in many applications, the disturbance is predicted and utilized to generate the control inputs.
Usually the disturbance is considered as a random variable, and the future disturbance is considered as a stochastic process.
There are a few ways to integrate the predicted stochastic disturbance information with the optimal control.
A common practice is to consider the disturbance as a Markov chain \cite{ripaccioli2010stochastic}\cite{sun2014velocity}\cite{zeng2015parallel}\cite{moura2010stochastic}. 
This representation can capture the first-order characteristics of the disturbance, but it is not capable of extracting long-term information on a larger time scale. 
Another way is to build a dynamic model for the stochastic system whose output is the disturbance. 
Then additional states describing the dynamics for disturbance are added to the existing model of the system for optimal control calculation. 
A disadvantage of this state augmentation method is the high computation if the disturbance is generated by a high-dimensional complex system.
Scenario trees can also be used to describe the predicted disturbance information \cite{de2005stochastic}\cite{di2013stochastic}\cite{meng2011scenario}. 
A tree structure is used to represent all the possible disturbance sequences in the future. 
The optimal control with a disturbance scenario tree can be computed using dynamic programming, which will be shown in this paper.

In general, to solve optimal control problems, methods like nonlinear programming \cite{borhan2011mpc}, quadratic programming \cite{rotenberg2010ultracapacitor}, and dynamic programming \cite{o2006dynamic}\cite{johannesson2007assessing} can be used.
When the future disturbance is described as a scenario tree, dynamic programming is usually the only practical method to compute the global optimal policy.
However, dynamic programming suffers from the curse of dimensionality, which means it can only be used to solve problems with a relatively small state space size.
For the optimal control problem with a disturbance scenario tree, the state space size will grow exponential as the size of the tree grows.
Therefore computing the predictive optimal control policy for a complicated scenario tree is very difficult.
Approximate dynamic programming can be applied to find a sub-optimal solution in many optimal control problems \cite{bertsekas2012approximate}. 
However a good approximated method for dynamic programming with scenario trees is not yet available.

In this work, we aim at finding the approximate global optimal control policy given a scenario tree of predicted future disturbance. 
We propose a novel clustering-based approximation of the scenario tree and design optimal predictive policies based on this approximation. 
Differently from the first-order Markov chain approximation, the scenario tree approximation can use information from the starting step to the current step. 
With the tree approximation, we can find the global optimal policy via dynamic programming with reduced computational load. 
Our work can also be embedded in the receding horizon framework to solve the optimal control problem in one step of model predictive control.

The rest of this paper is organized as follows. 
The problem formulation is given in Section II. 
The exact scenario tree solution and a clustering-based approximation solution are proposed in Section III and Section IV, respectively. 
Section IV provides numerical examples. 
We conclude the paper in Section VI.

\section{Problem Formulation}

\subsection{Predictive Optimal Control with Disturbance}

Consider the discrete-time dynamic system
\begin{equation}\label{eq1}
    x_{k+1} = f_k(x_k, u_k, w_k),   
\end{equation}
where the system state $x_k \in X \subseteq{\mathbb{R}^n}$, the control input $ u_k \in U(x_k) \subseteq{\mathbb{R}^m}$, the generalized disturbance $w_k\in{\mathbb{R}^q}$, and finite discrete time step $k \in K =\{0,1,...,N-1\}$. The initial condition $x_0$ is known.

The system has a running cost function
\begin{equation}\label{eq2}
    l_k = g_k(x_k, u_k, w_k)   
\end{equation}
and a terminal cost function
\begin{equation}\label{eq3}
    l_N = g_N(x_N),
\end{equation}
where $l_k \in \mathbb{R}$ and $l_N \in \mathbb{R}$.
The total cost is the sum of the running cost at each step and the terminal cost,
\begin{equation}\label{eq4}
    l =  \sum_{k=0}^{N-1}l_k + l_N.
\end{equation}

The goal is to find control policies $\pi_k$ which minimize the total cost $l$. 
The control policies need to be causal, which means that they can only use current and past information. 
It is assumed that the values of $x_k$ and $w_k$ are known at step $k$, then
\begin{equation}\label{eq5}
    u_k = \pi_k(x_0,...,x_k,w_0,...,w_k).
\end{equation}

If $w_0,..., w_{N-1}$ are known at step $0$ (future values of the disturbance are known), this optimal control problem can be easily solved via dynamic programming. 
In most applications, knowing the future values of the disturbance is not possible. 
Many predictive control applications use predicted nominal values of the future disturbance $\hat{w}_0, ...,\hat{w}_{N-1}$ to solve the optimal control problem. 
With the predicted nominal disturbance, the nominal optimal policy can be obtained using dynamic programming backward induction,
\begin{equation}
    J_k(x) = \min\limits_{u\in U(x)}\{g_k(x,u,\hat w_k) + J_{k+1}[f_k(x,u,\hat w_k)]\},
\end{equation}
\begin{equation}
    u_k(x) = \argmin\limits_{u\in U(x)}\{g_k(x,u,\hat w_k) + J_{k+1}[f_k(x,u,\hat w_k)]\}.
\end{equation}

\subsection{Disturbance Sequences}
In many cases, the generalized disturbance may be considered as the output of a very complicated stochastic system. 
For example, the driver's power demand is a reaction to the changing road profiles, traffic conditions, traffic lights, and pedestrians. 
In these cases, it is desired to develop a causal optimal controller with respect to not just the nominal disturbance, but all possible disturbance scenarios in the future.
In other words, we want to make a prediction that includes all possible future scenarios with their probabilities, and find the optimal control with respect to this stochastic future.

We define a disturbance sequence as $\bm{w} = [w_0 , ..., w_{N-1}]\in \mathcal{W}$, where $\mathcal{W}$ is the set of all possible disturbance sequences.
We use $\bm{w}(k)$ to denote the $k$-th element of this disturbance sequence, that is, $w_k$.
For each possible disturbance sequence $\bm{w}_i$ , its a priori probability is $p(\bm{w}_i)$.
It is assumed that all possible disturbance sequences $\mathcal{W}$ and the probabilities $p(\bm{w}_i)$ of each $\bm{w}_i\subseteq \mathcal{W}$ are predicted at step 0.
At a future step $k$, no additional information about $\bm{w}$ and the probabilities of $\bm{w}$ is provided except the value of $w_k$.
We use $\mathcal{W}_a$ to denote the set containing all disturbance sequences with non-zero probabilities at step 0.

\section{Optimal Control with Scenario Trees}

\subsection{Scenario Trees}

We can use a tree structure to represent all disturbance sequences in the set $\mathcal{W}_a$.
Suppose at step $k$, the disturbance sub-sequence that has been observed from step 0 to $k$ is $[w_0, w_1,...,w_k]$.
The information from this sub-sequence can be interpreted as that the full disturbance sequence $\bm{w}$ must satisfy $\bm{w}(0) = w_0$, ..., $\bm{w}(k) = w_k$.
If we define a set function $I(\cdot)$ on disturbance sub-sequences
\begin{equation} \label{eq8}
    \begin{aligned}
       &I_{k}([w_0,...,w_k]) \\
       =& \{\bm{w}\in \mathcal{W}_a|\bm{w}(0)=w_0,...,\bm{w}(k)=w_k\}. 
    \end{aligned}
\end{equation}
then given disturbance observation $[w_0, w_1,...,w_k]$, we can define a set $\mathcal{N}_k$ as $\mathcal{N}_k = I_{k}([w_0,...,w_k])$ to represent the information we have learned from the disturbance history.
We call $I_{k}([w_0,...,w_k])$ the admissible disturbance sequence set at step $k$.

At every step $k$, based on the observation of the current and historical values of $w$, we can determine $\mathcal{N}_k$. 
It is not difficult to see that $\mathcal{N}_k$ satisfies
\begin{equation} \label{eq_N_set}
  \mathcal{W}_a \supseteq \mathcal{N}_0 \supseteq \mathcal{N}_1 \supseteq ... \supseteq{N}_{N-1} = \{\bm{w}\},
\end{equation}
where $\bm{w}$ is the full disturbance sequence observed at step $N-1$.

The stochastic dynamic evolution of the disturbance sequence set $\mathcal{N}$ from step 0 to $N-1$ can be obtained when an initial value of $\mathcal{N}$ is given.
Before step 0, the only information about future disturbance is that $\bm{w} \in \mathcal{W}_a$, which means is $\mathcal{W}_a$ is the initial condition for $\mathcal{N}$.
Due to the property shown in (\ref{eq_N_set}), it is intuitive to use a tree structure to represent these dynamics.
This can be called a scenario tree.

\begin{figure}[ht!]
    \centering
    \includegraphics[scale=0.3]{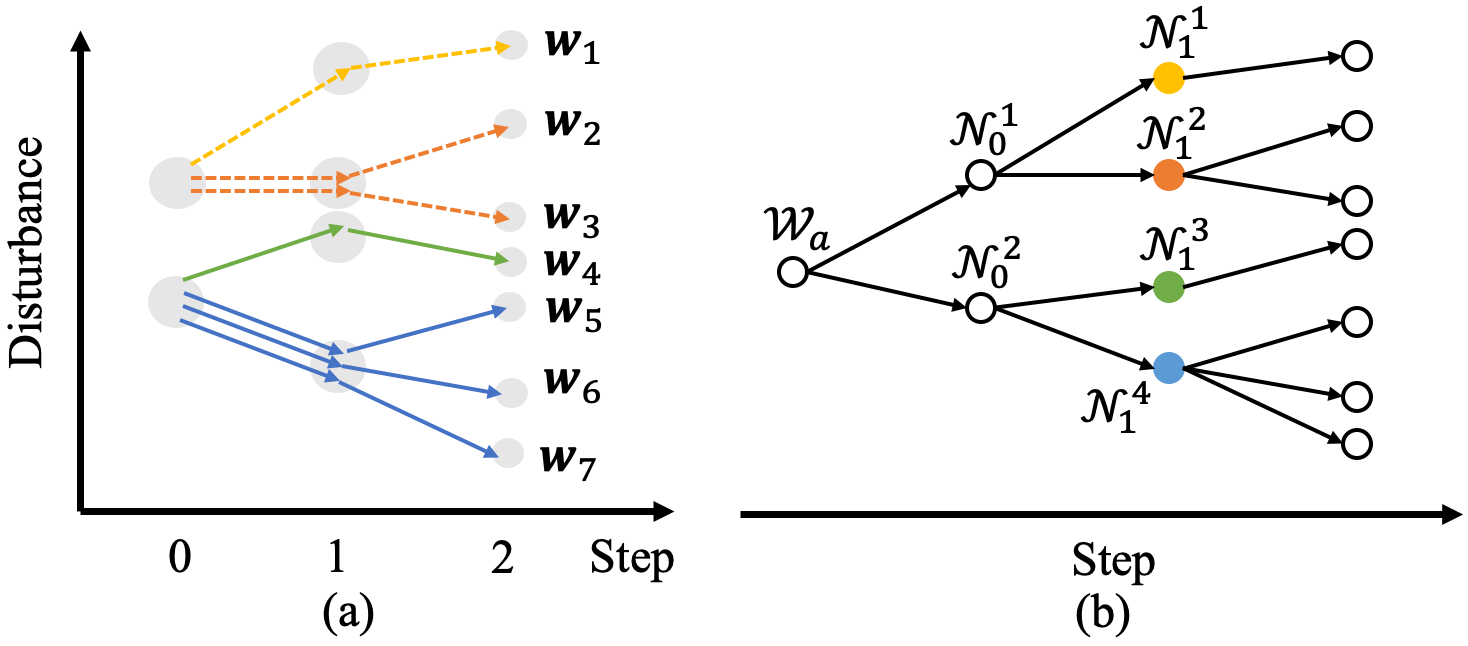}
    \caption{A scenario tree example: (a) 7 disturbance sequences; (b) the scenario tree representation of the 7 disturbance sequences.}
    \label{fig:tree_nodes}
\end{figure}

Fig. \ref{fig:tree_nodes} provides an example of a scenario tree.
There are 2, 4, and 7 possible $w$ values at step 0, 1, and 2, respectively, and there are 7 possible disturbance sequences.
In Fig. \ref{fig:tree_nodes} (b), each node of the tree is essentially a set of disturbance sequences $\mathcal{N}^i_k\subseteq \mathcal{W}_a$.
The root node, which is the initial set $\mathcal{W}_a$, contains all 7 sequences.
$\mathcal{N}_0^1$ contains the 3 dash-line sequences and $\mathcal{N}_0^2$ contains the 4 solid-line sequences.
At step 1, the four colored nodes $\mathcal{N}_1^i$ contain the disturbance sequences of the corresponding color. 
At step $k$, the node $\mathcal{N}_k$ can be considered as a measured state of the system, as $w_0, ...,w_k$ are known at step $k$.
The arrows in Fig. \ref{fig:tree_nodes} (b) can be considered as the possible state transitions of the tree node $\mathcal{N}$.

Define $p(\mathcal{N}_k)$ as the probability of the node $\mathcal{N}_k$, which is the sum of the probabilities of all disturbance sequences in this set,
\begin{equation} \label{eq9}
    p(\mathcal{N}_k) = \sum_{\bm{w}_i\in \mathcal{N}_k}^ip(\bm{w}_i).
\end{equation}

A node $\mathcal{N}_k$ may have multiple branches leading to multiple child nodes at step $k+1$. The conditional probability of the node $\mathcal{N}_k$ moving to its child node $\mathcal{N}_{k+1}$ can be obtained using the probabilities of the disturbance sequences in these two node sets,
\begin{equation}
    \begin{aligned}
        P[\mathcal{N}_{k+1} = I_{k+1}(w_0,...,w_k,w_{k+1})|\mathcal{N}_k = I_k(w_0,...,w_k)]  \\
        = p(I_{k+1}(w_0,...,w_k,w_{k+1}))/p(I_k(w_0,...,w_k))
    \end{aligned}
\end{equation}
In an exact scenario tree, $\mathcal{N}_k$ has stochastic dynamics in the form of a Markov chain as follows,
\begin{equation}
    P(\mathcal{N}^j_{k+1}|\mathcal{N}^i_k) = 
    \begin{cases}
        p(\mathcal{N}^j_{k+1})/p({\mathcal{N}^i_k}),& \text{if} \ \mathcal{N}^j_{k+1} \subseteq \mathcal{N}^i_k,\\
        0,                & \text{otherwise.}
    \end{cases}
\end{equation}

\subsection{Integration in Optimal Control}

Considering the system dynamics,
\begin{equation} \label{eq12}
    x_{k+1} = f_{\mathcal{N},k}(x_k,u_k,\mathcal{N}_k) = f_k[x_k,u_k,w_k(\mathcal{N}_k)],
\end{equation}
where $w_k(\mathcal{N}_k)=\bm{w}(k)$, $\forall{\bm{w}\in\mathcal{N}_k}$.
As $w_0,...,w_k$ is known,  $\mathcal{N}_k$ can be viewed as a measured state of the system.
The feedback control policies $\pi_k$ will be,
\begin{equation} \label{eq13}
    u_k = \pi_k(x_0,...,x_k,w_0,...,w_k) = \pi_k(x_k,\mathcal{N}_k).
\end{equation}
Using the pair $(x,\mathcal{N})$ as the new state, the predictive optimal control policy $\bm{\pi}^*$ can be obtained via dynamic programming backward induction.



One disadvantage of this scenario tree method is the exponential growing computational consumption with the number of steps and branches of the tree. 
In practice, it is common that a large number of possible future disturbance sequences are predicted, that is, $|\mathcal{W}_a|$ can be indeed very large.
To reduce the optimal policy computation load for a large number of disturbance sequences, we propose an approximation method for the scenario tree in the next section.

\section{Optimal Control with Scenario Tree Approximation}

\subsection{Scenario Tree Approximation}

In this section, we propose a clustering-based approximation method for the scenario tree. 
Instead of determining the tree node $\mathcal{N}_k$ using $I_k([\bm{w}(0), ..., \bm{w}(k)])$, a different way to map a disturbance sub-sequence $[w_0, w_1,...,w_k]$ to a node is used. 

At each step $k$, a clustering function $C_k$ is used to cluster disturbance sequences $\bm{w}$ with similar sub-sequences $[\bm{w}(0), ..., \bm{w}(k)]$. 
Clustering means classifying samples in the data set to a smaller number of clusters following specific rules that measure the similarity. 
Given any disturbance sub-sequence $[w_0 , ..., w_k]$, $C_k$  maps this sequence to an integer class number $i$, $i \in 1, 2, ..., n_k$ , where $n_k$ is the total number of clusters at step $k$. 

Then in the scenario tree approximation, elements in one node is defined by the following set,
\begin{equation}
    \mathcal{A}_k(i) = \{\bm{w} \in \mathcal{W}_a | C_k ([\bm{w}(0), ..., \bm{w}(k)]) = i\}.
\end{equation}
All disturbance sequences in $\mathcal{N}^i_k = \mathcal{A}_k(i)$ have their sub-sequences before step $k$ belonging to a same cluster under $C_k$.

\begin{figure}[ht!]
    \centering
    \includegraphics[scale=0.3]{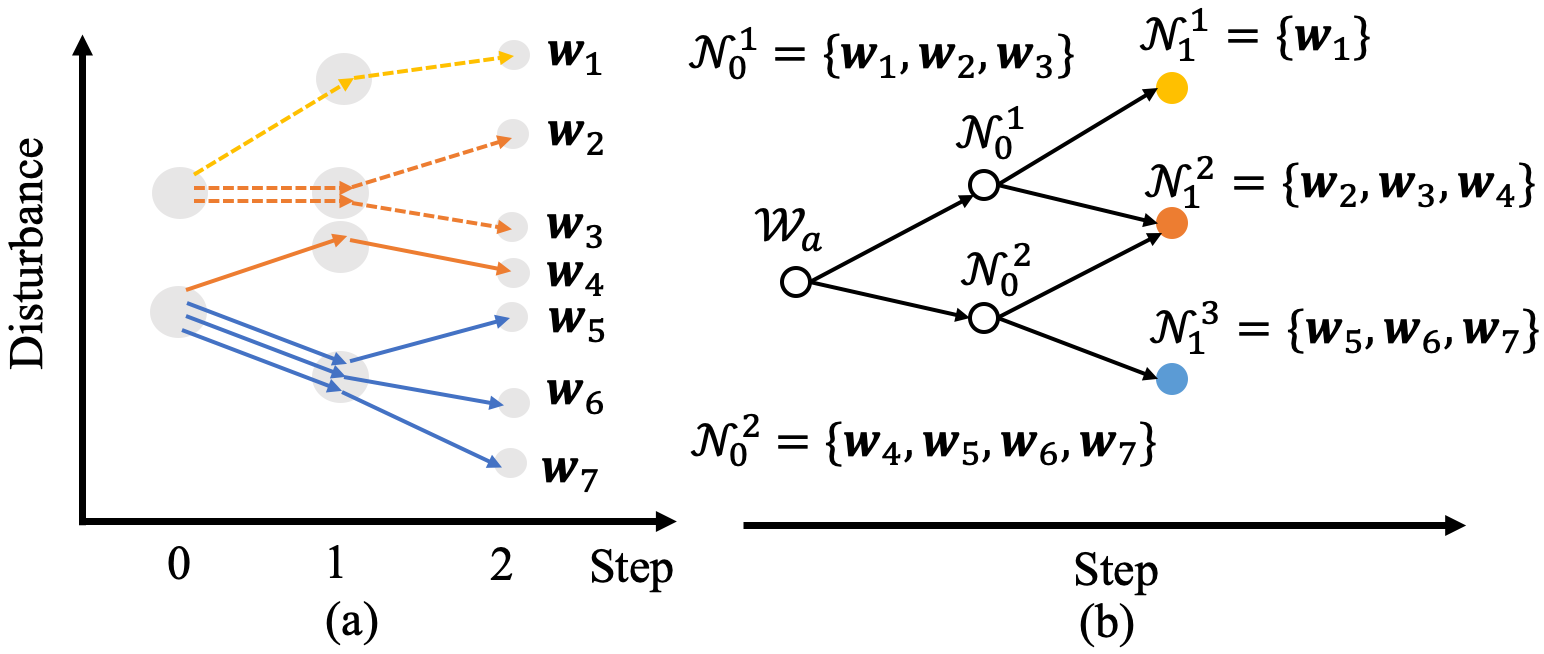}
    \caption{A clustering-based scenario tree approximation example: (a) 7 disturbance sequences; (b) the 3-cluster approximation of the scenario tree at step 2}
    \label{fig:cluster}
\end{figure}

An example that illustrates the clustering of the scenario tree is shown in Fig. \ref{fig:cluster}. 
We assume $n_k=3$, that is, we want to use no more than 3 clusters at each step to approximate the tree structure.
The clustering at step 0 is the same as the exact scenario tree, as the exact tree has only 2 nodes at step 0.
At step 1, a clustering based on $[\bm{w}(0),\bm{w}(1)]$ is performed, which generates 3 clusters: $\mathcal{N}_1^1$, $\mathcal{N}_1^2$, and $\mathcal{N}_1^3$.
After the clustering, the property in (\ref{eq_N_set}) no longer holds.
Therefore $\mathcal{N}$ is no longer in a tree structure as one child node $\mathcal{N}_{k+1}$ may have multiple parent nodes $\mathcal{N}_k$. 
However, $\mathcal{N}_k$ is still a measured state and its stochastic state transition is an approximation of the exact scenario tree. 

In the scenario tree approximation, $\mathcal{N}_k$ has the stochastic dynamics as follows,
\begin{equation}\label{eq_cluster_N_P}
    P(\mathcal{N}^j_{k+1} | {\mathcal{N}^i_k}) = p(\mathcal{N}^j_{k+1} \cap \mathcal{N}^i_k)/p(\mathcal{N}^i_k)
\end{equation}


The clustering-based approximation cuts down the number of states that representing disturbance from the number of all possible sub-sequences  $[w_0,..., w_k ]$ to a preset number $n_k$. 
This significantly reduces the state space size for backward induction in dynamic programming. 
As the number of clusters $n_k$ increases, the approximation becomes closer to the exact scenario tree. 
If the total sub-sequences number is large, we can even further reduce the computational load by random sampling in $\mathcal{A}_k(i)$ to approximate the expected cost and transition probabilities for $\mathcal{A}_k(i)$.

\subsection{A Two-Level Control Architecture}
One problem after the clustering approximation is that, $\bm{w}(k)$ may take different values in the same $\mathcal{N}_k$.
This means that at step $k$ the selection of a common control input $u_k$ that works for all $\bm{w} \in \mathcal{N}_k$ may be limited.
In an extreme case, such a common $u_k$ may not exist.
We design an extra mapping layer to guarantee that the control policy can offer admissible control input to the system. 
An admissible control input $u_k$ should guarantee that $x_{k+1} \in X$, where $x_{k+1} = f_k(x_k, u_k, w_k)$.
Instead of finding a feedback control policy $u_k = \pi_k(x_k,\mathcal{N}_k)$, we find a higher-level control 
\begin{equation}
    v_k = \pi_k(x_k,\mathcal{N}_k),
\end{equation}
and determine the lower-level control by 
\begin{equation}
\begin{aligned}
 u_k = G_u(x_k,v_k,w_k) =  \argmin\limits_{\text{admissible }u_k}(||v_k-u_k||).
 \end{aligned}
\end{equation}
Essentially the feedback control policy becomes $u_k=G_u[x_k,\pi_k(x_k,\mathcal{N}_k),w_k]=\pi'_k(x_k,\mathcal{N}_k,w_k)$. 
Because $G_u(\cdot)$ only depends on the values of $x_k,v_k,w_k$, we can define the mapping $G_u(\cdot)$ before the design of optimal control, and consider it as part of the stochastic system dynamics.

\subsection{Integration in Optimal Control}
In the cluster approximation, given $\mathcal{N}_k$, $w(\mathcal{N}_k) = \bm{w}(k)$, $\bm{w} \in \mathcal{N}_k$ is a random variable.
Therefore the system dynamics equation
\begin{equation}
\begin{aligned}
    f_{\mathcal{N},k}(x_k,u_k,\mathcal{N}_k) =& f_k[x_k,u_k,w_k(\mathcal{N}_k)] \\
=& f_k[x_k,G_u(x_k,v_k,w_k),w_k(\mathcal{N}_k)] \\
=&f_{\text{cluster},k}(x_k,v_k,\mathcal{N}_k)
\end{aligned}
\end{equation}
is a stochastic function whose output $x_{k+1}$ is a random variable.
The probability of $x_{k+1}$ and $\mathcal{N}_{k+1}$ can be determined if $x_k$, $\mathcal{N}_k$ and $v_k$ is given. 
The process of backward induction is similar to the exact scenario tree case, where the pair $(x,\mathcal{N})$ is considered as the state of a stochastic system.

\section{Numerical Examples and results}
\par In this section, we use two numerical examples to show the performance of the exact tree scenario and clustering-based approximation of the scenario tree. 

We consider the system,
\begin{equation}
    x_{k+1} = 0.9x_k + u_k -w_k,
\end{equation}
where $x$, $u$ and $w$ are scalars.
The system has a quadratic running cost
\begin{equation}
    g_k(x_k,u_k,w_k) = u^2_k,
\end{equation}
and a terminal cost
\begin{equation}
    g_N(x_N) = (-2-x_N)^2.
\end{equation}
The initial condition is $x_0 = 1.8$, and $N =10$. The constraints are $x \in [-2, 2]$ and $u \in [0, 1.6]$
In dynamic programming, $x$ and $u$ are quantized with an interval of 0.02.
Without loss of generality, we assume that every disturbance sequence in the predicted disturbance sequence set $\mathcal{W}_a$ has equal probabilities. 

\subsection{A Simple Scenario Tree}
\begin{figure}[ht!]
    \centering
    \includegraphics[scale=0.072]{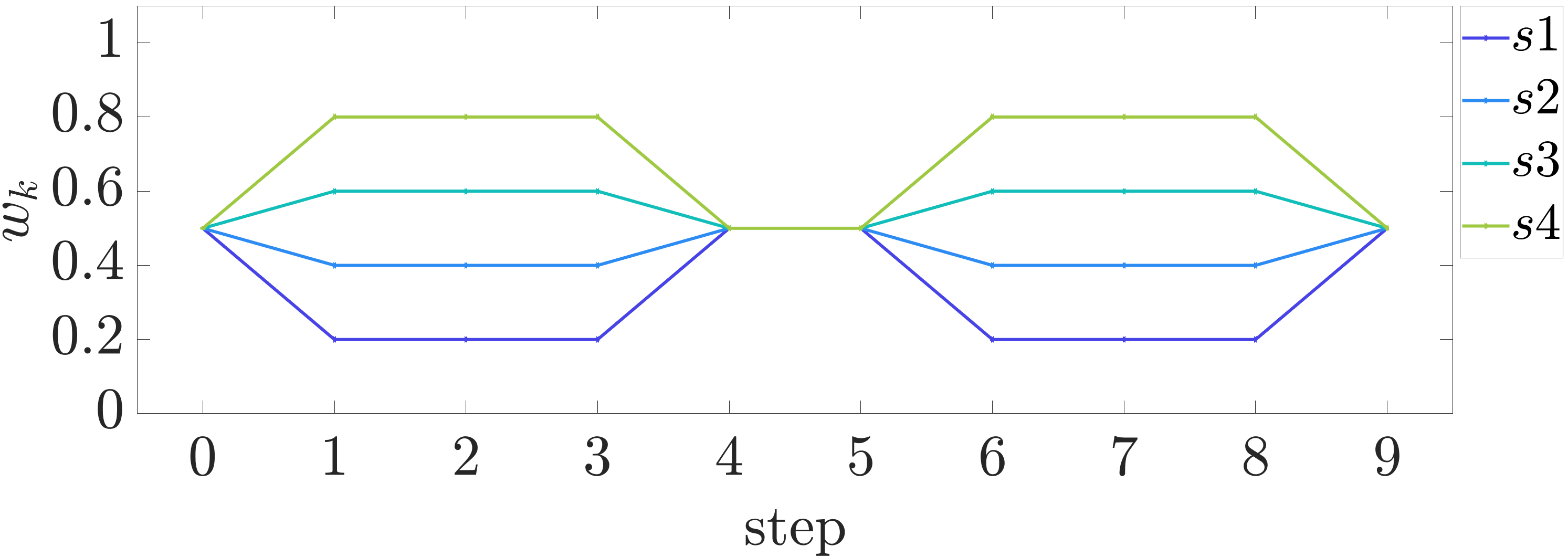}
    \caption{A simple disturbance scenario of 4 sequences. The 4 disturbance sequences are defined as s1, s2, s3, and s4.}
    \label{fig:simpleTree}
\end{figure}

A simple $\mathcal{W}_a$ with 4 disturbance sequences as shown in Fig. \ref{fig:simpleTree} is considered.
All sequences converge to a single point at step 4 and 5. 
We use this example to show why in some cases a scenario tree representation may be more appropriate than a non-stationary Markov chain representation.
The transition probability matrix $T_k$ of this non-stationary Markov chain at step $k$ can be defined as
\begin{equation}{\label{Markov_transition}}
    T_k(i,j) = P[w_{k+1} = w^j_{k+1} | w_k = w^i_{k}],
\end{equation}
for every possible $w^i_k$ and $w^j_{k+1}$. 
$T_k$ is computed using the disturbance sequence data for $k=0,1,...,N-1$. 

We use dynamic programming to compute two policies with two methods: (1) using the exact scenario tree to represent the future disturbance sequences and (2) using the non-stationary Markov chain to represent the future disturbance sequences.

The average cost for four disturbance sequences under the two methods is shown in Table \ref{tab:result1}. 
In this simple tree structure, the limitation of the Markov chain can be seen. 
The performance of the optimal predictive control with the scenario tree representation is better than that using the Markov chain.
This is due to the fact that the scenario tree keeps the multi-step long-term information of the disturbance sequences, while the Markov chain representation loses this information and only considers transitions between two steps.

\renewcommand{\tabcolsep}{1mm}
\begin{table}[htbp]
    \centering   
    \caption{Average cost results of the 4 disturbance sequences}
    \label{tab:result1}
    \begin{tabular}{c c c}
        \toprule
        Method & Average cost & Computation time\\
        \midrule
        Exact Scenario Tree  & 0.3180 & 5.52s \\          
        Non-Stationary Markov Chain & 0.3429 & 18.29s\\    
        \bottomrule
    \end{tabular}
\end{table}

\subsection{A Complex Scenario Tree with Approximation}

\begin{figure}[ht!]
    \centering
    \includegraphics[scale=0.072]{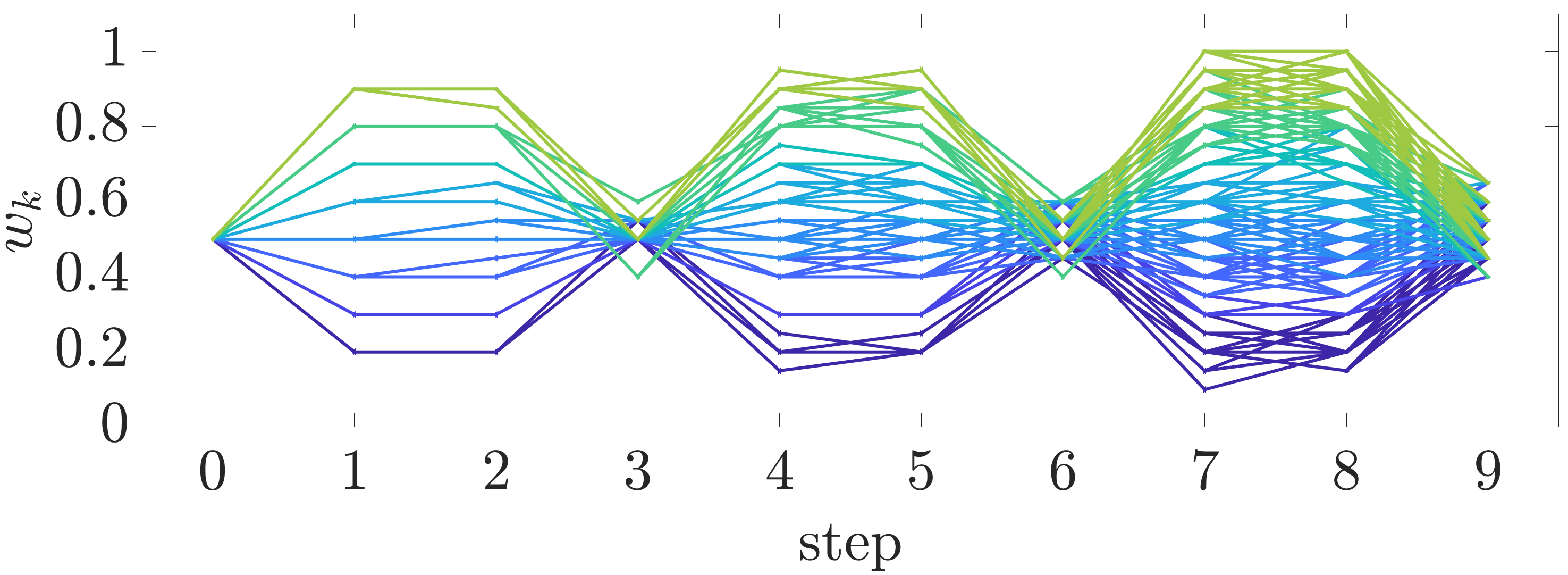}
    \caption{A complex disturbance scenario of 1447 sequences.}
    \label{fig:complexTree}
\end{figure}

In this example, we use a more complex scenario tree to demonstrate the performance of the clustering-based tree approximation method. 
1447 disturbance sequences are generated as shown in Fig. \ref{fig:complexTree}.

\begin{figure}[ht!]
    \centering
    \includegraphics[scale=0.16]{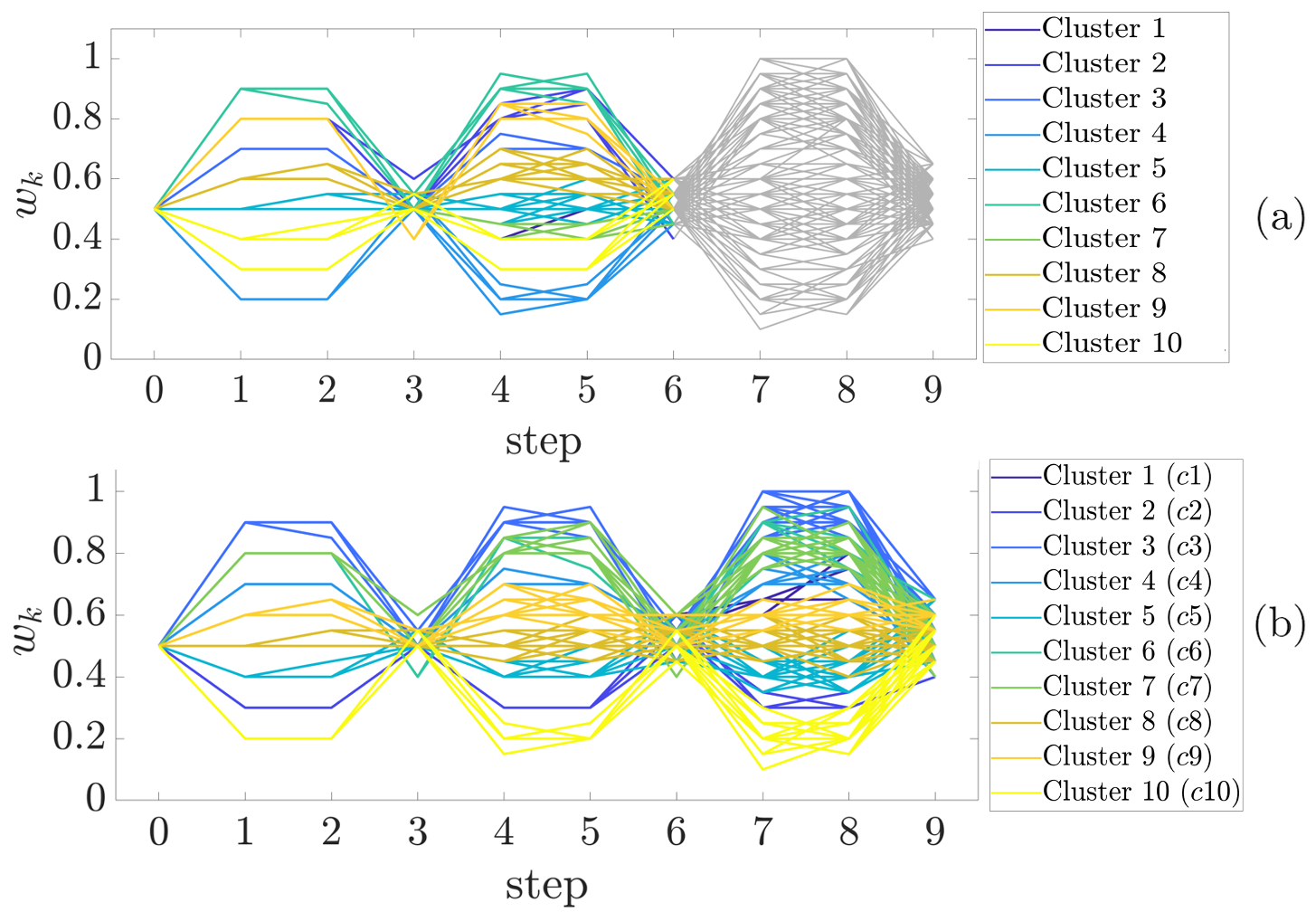}
    \caption{Clustering results for disturbance sequences at (a) step $6$,  and (b) step $9$.}
    \label{fig:clustering_result}
\end{figure}

We use dynamic programming to compute two policies with three methods: (1) using the exact scenario tree representation, (2) using the clustering-based scenario tree approximation representation, where $n_k=10$ (10 clusters for each step), and (3) using the non-stationary Markov chain to represent the future disturbance sequences.
In this example, we use the k-means clustering\cite{arthur2006k} as the function $C_k$ to cluster the disturbance sub-sequences. For k-means, we wish to choose $n_k$ centers $\mathscr{C}$ of the data to minimize the potential function, 
\begin{equation}
    \phi = \sum_{w \in W} \min\limits_{c\in \mathscr{C}}||w - c||^2,
\end{equation}
The clustering result examples at step $6$ and $9$ are shown in Fig. \ref{fig:clustering_result}.
In the non-stationary Markov chain method, to keep the size of the state space the same as the clustering-based method, we quantize the disturbance to a 10-value grid and obtain a time-dependent transition matrix using (\ref{Markov_transition}).

\renewcommand{\tabcolsep}{1mm}
\begin{table}[htbp]
    \centering    
    \caption{Average cost results of the 1447 disturbance sequences}
    \label{tab:result2}
    \begin{tabular}{c c c}
        \toprule
        Method & Average cost & Computation time\\
        \midrule 
        Exact Scenario Tree & 0.4407 &  487.90s \\
        Clustering-Based Approximation  &  0.4432 & 32.52s \\ 
        Non-Stationary Markov Chain & 0.4992 & 17.67s \\      
        \bottomrule
    \end{tabular}
\end{table}

\par
\begin{figure}[ht!]
    \centering
    \includegraphics[scale=0.24]{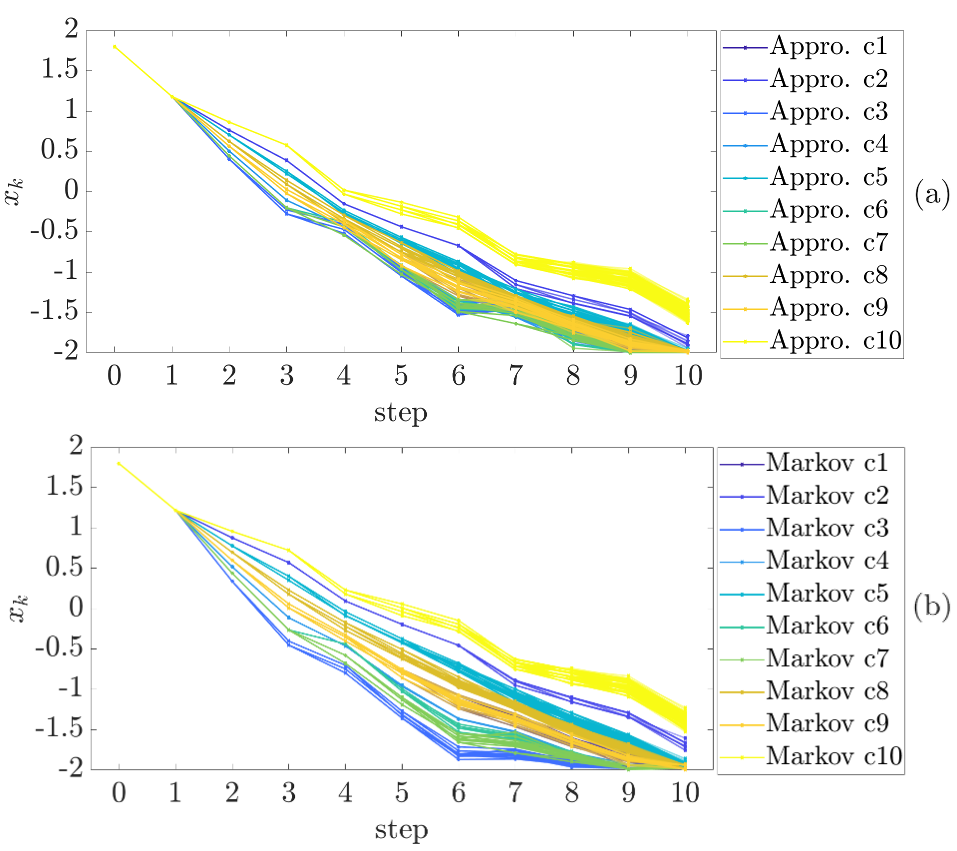}
    \caption{State trajectories for (a) the clustering-based scenario tree approximation method and (b) the non-stationary Markov chain method of the 1447 disturbance sequences. Each colored trajectory  is corresponding to a disturbance sequence in one cluster of step 9.}
     \label{fig:x_complex}
\end{figure}

 The average cost for all 1447 disturbance sequences is shown in Table \ref{tab:result2} and the state trajectories are shown in Fig. \ref{fig:x_complex}. 
For the computation time, all results are from codes written in MATLAB and run on a Mac laptop with Intel Core i7-9750H. 
To optimize computing efficiency for the clustering-based method, we build a static table that maps the discretized $x_k,u_k,w_k$ values to corresponding $x_{k+1}$ and $l_k$. 
It takes 6.3s to build the static map (included in the 32.52s computation time). The map can be reused for updating policies from different clustering approximation results unless the system dynamics or the cost function changes. The clustering-based method performs much better than the non-stationary Markov chain method and ensures a similar performance as the exact scenario tree method. However, the clustering approximation method significantly reduces computation time from the exact tree method. The clustering-based method captures more disturbance sequence features from previous steps while the Markov chain loses this information.

\section{CONCLUSIONS}

In this paper, a predictive optimal control method with a clustering-based scenario tree approximation of disturbance sequences is proposed. 
The method is demonstrated in numerical examples, where the proposed method performs better than the traditional method which considers the disturbance as a Markov chain.


\addtolength{\textheight}{-12cm}   



\bibliographystyle{IEEEtran}
\bibliography{Main}

\end{document}